\documentstyle[aps,prb,floats,epsf]{revtex}

\begin{document}
\draft
\twocolumn[\hsize\textwidth\columnwidth\hsize\csname@twocolumnfalse%
\endcsname
\title{Dynamical universality classes of
the superconducting phase transition}

\author{Jack~Lidmar$^a$, Mats~Wallin$^a$,
Carsten~Wengel$^b$, S.~M.~Girvin$^c$, and A.~P.~Young$^d$}
\address{$^a$Department of Theoretical Physics,
Royal Institute of Technology, SE-100 44 Stockholm, Sweden\\
$^b$ Institut f{\"u}r theoretische Physik, Universit{\"a}t
G{\"o}ttingen,
Bunsenstr.~9, 37073 G{\"o}ttingen, Germany \\
$^c$ Department of Physics, Indiana University, Bloomington, IN
47405, USA\\
$^d$ Department of Physics, University of California, Santa
Cruz, California 95064, USA
}
\date{\today}
\maketitle
\begin{abstract}
We present a finite temperature Monte Carlo study of the XY-model
in the vortex representation, and study its dynamical critical
behavior
in two limits. The first neglects
magnetic field fluctuations, corresponding to the absence of
screening, which should be a good approximation
in high  $T_c$  superconductors ($\kappa\to \infty$)
except extremely close to the critical point.  
Here, from finite size scaling of the linear resistivity
we find the dynamical critical exponent of the vortex motion to be
$z\approx 1.5$.  The second limit includes magnetic field
fluctuations in the strong screening limit ($\kappa\to 0$)
corresponding to the true asymptotic inverted XY critical regime, where
we find the unexpectedly large value $z\approx 2.7$. 
  We compare these 
results, obtained from dissipative dynamics in the vortex representation,
with the universality class of the corresponding model in the phase
representation with propagating (spin wave) modes.
We also discuss the effect of disorder and the
relevance of our results for experiments.
\end{abstract}
]

\section{Introduction}

Considerable confusion currently exists, both theoretically and
experimentally, regarding the dynamical universality class of the
zero-field superconducting phase transition in high temperature
superconductors.  The short coherence length in these materials leads
to large, non-gaussian fluctuations, and   
there is some experimental evidence that the static behavior is
that of the 3-dimensional XY
model\cite{ffh91,friesen-muzikar,%
3DXY1,3DXY2,3DXY3,moloni-prl,moloni-prb,ginsberg},
as in the lambda transition in superfluid helium.
The dynamical universality class of the lambda transition in
superfluid helium is that of a two component order parameter 
coupled to a conserved density which
gives a propagating mode (second sound in $^4$He; a spin wave in the
XY spin model) in the broken symmetry phase. This is model E in the
notation of Hohenberg and Halperin.\cite{hohenberg77}
However, the dynamical behavior of a superconductor could be 
very different, since the
lattice acts as a momentum sink destroying Galilean invariance making
the system more like helium in a porous medium.  The existence of
disorder does not destroy the Goldstone mode in the ordered phase but
it does affect the vortices and the normal-fluid quasiparticles which
tend to equilibrate with the lattice rather than co-moving with the
condensate.   

Furthermore, in d-wave superconductors, scalar disorder causes
pair-breaking and quasiparticle branch recombination which
may make it inappropriate to assume  particle number conservation
in the dynamics.  There does not appear to have been any work to date
on this question.  

A related issue is that of the role of the long-range Coulomb
interaction, which severely suppresses longitudinal current
fluctuations, leaving only the transverse currents associated with
vortices.  However it is also possible that the low superfluid
density and screening from the high normal fermion density will turn 
on the low energy longitudinal Carlson-Goldman 
fluctuations\cite{carlson-goldman}  
of the order parameter as $T_c$ is approached.
These microscopic fermionic effects  may further confuse the data 
analysis and affect the width of the critical regime.

The issues raised above remain largely unresolved.  
The particular issues that we will explicitly discuss here are the role of 
magnetic screening and the role of disorder.  
In an extreme type-II, system, coupling to 
gauge fluctuations is weak,\cite{ffh91} but nonetheless, in 
principle, magnetic screening becomes important extremely close to the
critical point where the system crosses over to inverted XY
behavior.\cite{herbut,dasgupta81,inverted}  
The static correlations related to 
this issue have recently been studied numerically by Olsson and 
Teitel.\cite{olsson-teitel}  Here we will address the dynamics.

Most theoretical studies of critical dynamics in XY like spin systems
have focussed on Landau-Ginsburg representations of the problem 
involving the {\em phase} (i.e. angle) of the spin.  For static
properties, there also exist equivalent dual
representations\cite{dasgupta81,villain75,jose77,kleinert89} in terms
of interacting {\em vortex} degrees of freedom. Although the static
properties of the phase and vortex representations are the same, 
there is no reason, {\em a priori}, why the dynamical universality 
classes should be the same.  

For the reasons discussed above it may be more appropriate, 
for superconductors, to consider a model with overdamped dissipative
dynamics of the topological defects (the vortices). 
This is the approach that we will take here.


To add dynamics to the the phase representation, one can either
include just dissipative dynamics, model A,\cite{hohenberg77} in
which case the dynamical exponent, $z$, is close to 2, or one can 
incorporate the propagating (spin wave) modes, model
E,\cite{hohenberg77} for which $z$ is exactly 3/2 in three dimensions,
($d/2$ in $d$-dimensions). For the vortex representation, which has
{\em discrete} variables, the natural dynamics is purely dissipative,
such as that generated by Monte Carlo simulations (in which Monte
Carlo time is equated with real time).  Naively, it would
seem unlikely that the {\em dissipative} dynamics of the vortex
representation would be in the same universality class as the dynamics
of model E (phase representation), which has {\em propagating} modes.
Surprisingly, recent results by Weber and Jensen\cite{weber97}
come to the opposite conclusion.  They find, for unscreened vortex 
interactions ($\kappa=\infty$) and overdamped dynamics, that the 
dynamical exponent is $z=d/2=1.5$, precisely the value one expects in
model E dynamics, and considerably {\em less} than the value generally 
found with dissipative dynamics ($z \approx 2$).

In this paper we present results of Monte Carlo calculations of the
dynamical critical exponents for the 3-dimensional XY model, in the 
vortex representation, with and without magnetic screening.  For no 
screening, we confirm the unexpected result of Weber and 
Jensen\cite{weber97}, while by contrast, for strong magnetic screening,
we find a rather large {\em enhancement} of $z$.
Although it is known from the Harris criterion and verified
numerically\cite{moon} that 
uncorrelated  
disorder is weakly irrelevant at
the 3-dimensional XY critical point, the effect of such disorder on the
dynamical properties is unknown.\cite{columnar}  
 We therefore also investigate the 
effects of disorder on the model with strong screening.

\section{The models}

The model under investigation here is the XY--model with a
fluctuating vector potential,
\begin{equation}
  \label{ham-phase}
  {\cal  H} = -J \sum_{\langle i,j \rangle} \cos(\phi_i - \phi_j
  -\lambda_0^{-1 } a_{ij}) + \frac{1}{2} \sum_\Box [{\bf \nabla}
  \times {\bf a}]^2,
\end{equation}
where $J$ is the coupling constant 
(set to unity in the simulations),    
 the $\phi_i$ denote the phases of the condensate on the sites $i$ of a
simple cubic lattice of size $N=L^3$ with periodic boundary
conditions. The sum is taken over all nearest neighbors
$\langle i,j \rangle$. Additionally, we have a fluctuating gauge
potential $a_{ij}$ with the gauge constraint
$[{\bf \nabla}\cdot {\bf a}]=0$, and $\lambda_0$ denotes the bare
screening length. The last term describes the magnetic energy, where
the sum runs over all elementary plaquettes on the lattice and the curl
is the directed sum of the gauge potential round a plaquette.

In our work, the main focus is equilibrium vortex
dynamics. However, vortices are only {\em implicitly}
present in the above model
through the relation
\begin{equation}
\oint \nabla \phi({\bf r})\cdot {\rm d} {\bf r}=2\pi n  ,
\end{equation}
where $n=0,\pm 1,\pm 2,...$ denotes the net vorticity encircled by
the integration contour.   
In order to analyze the critical behavior of this model due to vortex
fluctuations {\em explicitly}, it is easier to go from the above {\em
phase}
representation of the
XY--model to its {\em vortex} representation. This is achieved by
replacing the cosine in Eq.~(\ref{ham-phase})  with the periodic
Villain function and performing fairly standard
manipulations\cite{dasgupta81,villain75,jose77,kleinert89} to obtain
\begin{equation} \label{ham-vortex}
{\cal H}_V = \frac{1}{2} \sum_{i,j} {\bf n}_i \cdot {\bf n}_j\,
G_{ij}[\lambda_0].
\end{equation}
Here, the ${\bf n}_i$ are vortex variables which sit on the links of
the {\em dual lattice} (which is also a simple cubic lattice) and
$G_{ij}$ is the screened lattice Green's function
\begin{equation} \label{greens}
G_{ij}[\lambda_0] = J\frac{(2\pi)^2}{L^3} \sum_{\bf k} \frac{
\exp[{\rm i }\, {\bf k}\cdot({\bf r}_i - {\bf r}_j)]}
{2\sum_{m=1}^3 [1- \cos(k_m)] + \lambda_0^{-2}}.
\end{equation}
In the long range case, $\lambda_0 \to \infty$, the divergent ${\bf
k=0}$ contribution has to be excluded from the sum and the constraint
$\sum_i {\bf n}_i = {\bf 0}$ has to be imposed.  In the short range
case (finite $\lambda_0$) this is not necessary.
The transformations involved in going from the phase to the vortex
representation also yield the local constraints $[{\bf \nabla}\cdot
{\bf n}]_i=0$, i.e., there are no magnetic monopoles (zero divergence
constraint). It is important to note, that the vortex representation
does not contain spin wave degrees of freedom anymore, since they
have
been ``integrated out'' in the transformation procedure.

We will be interested in two limits of the above model: (i) no
screening ($\lambda_0\to \infty$), corresponding to the extreme
type--II limit ($\kappa\to \infty$), where the individual vortex
lines
have long range
interactions; (ii) strong screening ($\lambda_0\to 0$), i.e., short
range interactions, which is supposed to be the correct description
of a superconductor
extremely close to the critical point,\cite{ffh91} though the size of
the critical region where such 
screening is relevant may be too small to
be observable in practice.
In this limit the interaction reduces to $G_{ii} = J(2 \pi
\lambda_0)^2$ and $G_{i \ne j} = 0$ (plus exponentially small
corrections of order $\exp(-r/\lambda_0)$).  We will in this case use
units where $J(2 \pi \lambda_0)^2 = 1$.
The resulting Hamiltonian is
of the very simple form
\begin{equation} \label{ham-vortex-screen}
{\cal H}_V = \frac{1}{2} \sum_i {\bf n}_i\cdot{\bf n}_i,
 \quad (\lambda_0\to 0).
\end{equation}
Note, however, that this Hamiltonian is not trivial, since the
constraints on the local divergence effectively generate interactions
among the ${\bf n}_i$. Note further, that this is also the dual
representation of the XY model {\em without} screening, in which the
temperature scale is inverted.\cite{dasgupta81,kleinert89,wengel96}
The static universality class of
Eq.~(\ref{ham-vortex-screen}) is 
then the same as that of
Eq.~(\ref{ham-vortex}) with $\lambda_0=\infty$,
and is given by XY exponents.\cite{yeomans}
The dynamical universality class, however, may be
different and determining it is one of the goals   
 of our study.

In addition to the pure short-range model, we also study the
short-range model with quenched random local $T_c$. In this case
we replace Eq.~(\ref{ham-vortex-screen}) by
\begin{equation} \label{ham-vortex-screen-disorder}
{\cal H}_V = \frac{1}{2} \sum_{i,\mu} \xi_{i\mu}  n_{i\mu}^2, \quad
(\lambda_0\to
0,\; \xi_{i\mu}\ \mbox{random}).
\end{equation}
where $\xi_{i\mu}$ is uniformly distributed in the interval $[0.5,1.5]$.

\section{Monte Carlo simulation and finite size scaling}

We simulate the following model Hamiltonians in the vortex
representation:
\begin{enumerate}
\item
Eq.~(\ref{ham-vortex}) with $\lambda_0\to \infty$
\item
Eqs.~(\ref{ham-vortex-screen}) and
(\ref{ham-vortex-screen-disorder}),
which corresponds to $\lambda_0\to 0$.
\end{enumerate}
We take simple cubic lattices of size $L^3$ where $4 \le L \le $12--64.
Periodic boundary conditions are imposed. We start with
configurations
with all ${\bf n}_i = {\bf 0}$, which clearly satisfies the
constraints,
and a Monte Carlo (MC) move consists of trying to create a closed
vortex loop
around a plaquette.\cite{global}
This trial state is accepted according to the heat bath
algorithm with probability $1/[1+\exp(\beta \Delta E)]$,
where $\Delta E$ is the change of energy and $\beta=1/T$.

Each time a loop is formed it generates a voltage pulse $\Delta Q =
\pm 1$ perpendicular to its plane, the sign depending on the
orientation of the loop.
This leads to a net electric field\cite{hyman95}
\begin{equation}
E(t) = \frac{h}{2e} J^V(t)
\quad \mbox{with} \quad
J^V(t) =  \frac{\Delta Q}{\Delta t},
\label{voltage}
\end{equation}
where $J^V(t)$ is the vortex current density, and $\Delta t = 1$ for
one full sweep through the system, where, on average, an attempt is
made to create or destroy one vortex loop per plaquette.

The nonlinear I--V characteristics of the inverted XY
model can be modeled as the electric field $E$, due to vortex current
response in the presence of a uniform Lorentz force on the vortex
lines, proportional to the applied current density $J$.\cite{caveat1}
In addition, the linear response resistance can be calculated from the
equilibrium voltage--voltage fluctuations via the Kubo
formula\cite{young94}
\begin{equation} \label{kubo}
R = \frac{1}{2T} \sum_{t=-\infty}^{\infty} \Delta t
\langle V(t)V(0) \rangle
\end{equation}
Here, $\langle \cdots \rangle$ denotes the thermal average, and the
voltage across the sample is $V(t) = L E(t)$.  The resistivity is
$\rho = L^{d-2}R$.

Since we are working with lattices of finite length $L$, one has to
employ finite size scaling techniques to extract the critical
behavior. A detailed scaling theory has been developed for
superconductors by Fisher et al.\cite{ffh91} and we now summarize
the results from it which will be needed for our data analysis.

Near a second order phase transition the linear resistivity obeys the
scaling law
\begin{equation}
\rho_{\rm lin}(T,L)=L^{-(2-d+z)}\, \tilde{\rho}\left[
L^{1/\nu}(T-T_c)
\right ],
\label{rho_lin_scale}
\end{equation}
where $\nu$ is the correlation length exponent,
$z$ is the dynamical exponent and $\tilde{\rho}$ is a scaling
function.
At the critical temperature, $\tilde{\rho}(0)$ becomes a
constant and therefore $\rho_{\rm lin}(T_c,L) \sim L^{-(2-d+z)}$.
If we plot the ratio of $\rho_{\rm lin}$ for different system sizes
against $T$, then
\begin{equation} \label{intersect}
\frac{\ln[\rho_{\rm lin}(L)/\rho_{\rm lin}(L^{\prime})]}
{\ln[L/L^{\prime}]} = d-2-z \quad\mbox{at}\;\;  T_c,
\end{equation}
i.e., all curves for different pairs $(L,L^{\prime})$ should
intersect and one can read off the values of $T_c$ and $z$. We will
refer to this kind of data plot as the {\em intersection method}.
With the values of $T_c$ and $z$ determined by the intersection
method
we can then use a scaling plot according to Eq.~(\ref{rho_lin_scale})
to obtain the value of $\nu$.

A similar analysis\cite{ffh91,columnar} shows that, above a
characteristic current scale
$J_{\rm nl}$, which varies as
$\sim L^{-2}$ 
at the critical point, non-linear response
 sets in and the electric field varies as
\begin{equation}
E\sim J^{(1+z)/2}.
\label{Enonlin}
\end{equation}
    
It is useful to locate the critical temperature from equilibrium
properties instead of the dynamic scaling of the linear resistivity,
since such simulations are easier to converge.  In the short-range
case we do this by considering an ensemble with a fluctuating winding
number $W$, defined by
\begin{equation}
W_\mu = \frac{1}{L} \sum_i { n}_{i\mu}.
\end{equation}
{}From the finite size scaling relation
\begin{equation}
\left< W_\mu^2 \right> = f(L^{1/\nu}(T-T_c))
\end{equation}
we can locate the critical point in the short range case, both with and
without disorder using the fact that the winding number is scale
invariant at the critical point.
 Note that the simulations of this quantity require
global moves, where vortex lines going all the way through the system
are created and destroyed, and therefore do not represent the dynamics
of the system in a realistic way.  This does not pose a problem since
we use this calculation solely as a way to accurately locate the
critical point and not to follow the dynamics.

Because different winding number classes are difficult to equilibrate
efficiently in large systems in this model, 
 we use an (exact)  duality transformation from the short-range
vortex models, Eqs.~(\ref{ham-vortex-screen}) and
(\ref{ham-vortex-screen-disorder}),
back to an XY phase model with a Villain
interaction:\cite{villain75,dasgupta81}
\begin{eqnarray} \label{villain}
Z_V&=&\sum_{\{{\bf n}_{i\mu}\}} \delta_{\nabla\cdot {\bf n}_i,0}
e^{-\frac{1}{2T}\sum_{i\mu}\xi_{i\mu}{\bf n}_{i\mu}^2 } \nonumber \\
&=& \int\limits_0^{2\pi} \left[ \prod_i d\theta_i \right]
\sum_{\{{ m}_{i\mu}\}} e^{-\sum_{i\mu}{T \over 2\xi_{i\mu}} \left(
\theta_i-\theta_{i+{\bf e}_\mu} -2\pi { m}_{i\mu}\right)^2}
\end{eqnarray}
(a constant prefactor was suppressed in the last equality).  
Performing the sum over the integer dummy variables ${m}_{i\mu}$ leaves a
phase-only model.  This phase representation allows us
to take advantage of the Wolff algorithm in the
simulation which largely circumvents problems of critical slowing
down and equilibration in the limit of large system sizes.\cite{wolff} 
The spin-wave stiffness of the dual model is
given by $\rho_s = \partial^2 f / \partial
\tilde{A}_\mu^2|_{\tilde{A}=0}$, where $\tilde{A}_\mu$ is a constant
vector potential
added to the phase gradient in Eq.\ (\ref{villain}).  This is related
to the winding number fluctuations in the vortex model by $\rho_s = T
L^{2-d} \left< W_\mu^2 \right>$.

\section{Data analysis}

In this section we analyze our simulation data for the XY-model
in the vortex representation, starting with the model in
Eq.~(\ref{ham-vortex}) with long range interactions, i.e., neglecting
screening ($\lambda_0\to \infty$).

\subsection{Long range interactions}

\begin{figure}[htb]
\centerline{\epsfxsize 7cm\epsfbox{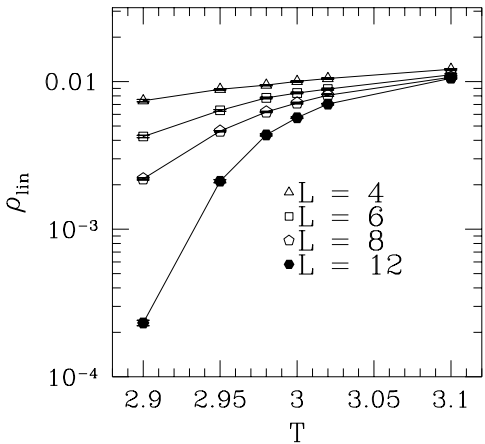}}
\caption{Linear resistivity $\rho_{\rm lin}$ vs.\ $T$ for the vortex
  model of Eq.~(\protect\ref{ham-vortex}) without screening,
  i.e. $\lambda_0 = \infty$.}
\label{purel_pic}
\end{figure}

In Fig.~(\ref{purel_pic}) we show the data for the linear resistivity
$\rho_{\rm lin}$ plotted vs.\ $T$ obtained from a simulation of the
Hamiltonian~(\ref{ham-vortex}) with $\lambda_0=\infty$.  One observes
that at high temperatures ($T=3.1$) there is hardly any size
dependence in the data, consistent with a correlation length which is
smaller than the system size. As one goes to lower temperature, the
size dependence becomes stronger. This indicates critical behavior,
since in the limit $L\to\infty$ the linear resistivity should go to
zero below $T_c$.

\begin{figure}[htb]
\centerline{\epsfxsize 7cm\epsfbox{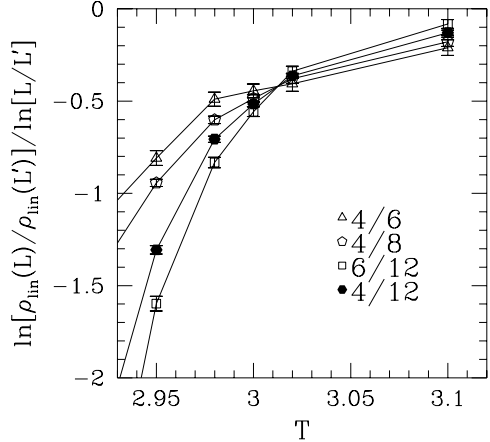}}
\caption{Linear resistivity, plotted according to the {\it
intersection
    method}, vs.\ $T$ with the data of
Fig.~(\protect\ref{purel_pic}). At the
    intersection one can
    read off $T_c\approx 3.01$ and $y \approx -0.45$,
    corresponding to $z\approx 1.45 (\pm 0.05)$.}  
\label{purel-is_pic}
\end{figure}

However, since it is
impossible to locate the critical temperature by this kind of plot,
we show, in Fig.~(\ref{purel-is_pic}),
the same data of Fig.~(\ref{purel_pic}) plotted according
to the intersection method.
The curves intersect at approximately $T_c=3.01\, (\pm 0.01)$ 
and $y\approx -0.45$ corresponding to $z=1.45\, (\pm 0.05)$,
very similar to the result of
Weber and Jensen, who find $z=1.51\, (\pm 0.03)$.\cite{weber97}
Also, the value of $T_c$
agrees nicely with the one obtained earlier from simulations by
Dasgupta and Halperin.\cite{dasgupta81}

\begin{figure}[htb]
  \centerline{\epsfxsize 7cm\epsfbox{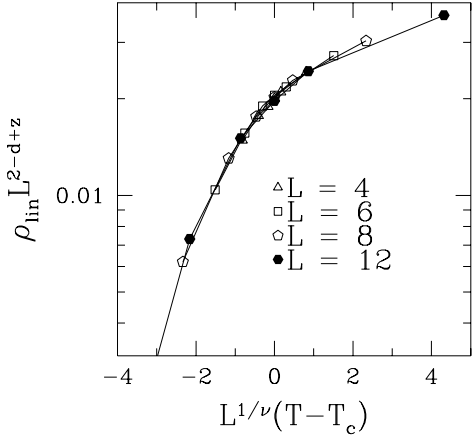}}
\caption{Scaling plot of $\rho_{\rm lin}$ according to
  Eq.~(\protect\ref{rho_lin_scale}). Best scaling was achieved
  with $T_c=3.01\,(\pm 0.01)$, $\nu=0.66\,
  (\pm 0.01)$, and $z=1.5\,(\pm 0.05)$.}
\label{purel-scale_pic}
\end{figure}

Having established $T_c$ and
$z$ we can now perform a scaling plot of the data according to
Eq.~(\ref{rho_lin_scale}). In Fig.~(\ref{purel-scale_pic})  we plot
$\rho_{\rm lin} L^{2-d+z}$ vs.\ $L^{1/\nu}\,(T - T_c)$ and find that
the data collapses best with
$T_c=3.01\pm 0.01$, $z=1.5 \pm 0.05$ and $\nu=0.66 \pm0.01$.
This independent result confirms that the dynamical scaling ansatz for
$\rho_{\rm lin}$ yields the expected value
of the correlation length exponent $\nu$ as well as a consistent value
for $z$. 

\subsection{Short-Range Interactions}

In this section we will first describe how we locate the critical
point of the pure and disordered short range models defined by
Eqs.~(\ref{ham-vortex-screen}) and
(\ref{ham-vortex-screen-disorder}),
using finite size scaling analysis of Monte Carlo data for static
quantities.  Then, using dynamic scaling analysis we determine the
dynamical exponents both from equilibrium vortex dynamics
simulations,
and from driven nonequilibrium simulations.

In our simulations of Eq.~(\ref{ham-vortex-screen}), we used
$10^6-10^7$ sweeps to calculate averages, and discard the initial
$10\%$ of the data for equilibration.  For the disordered case we
average over $10-100$ samples to obtain small fluctuations.

\begin{figure}[htb]
\epsfxsize=\columnwidth\epsfbox{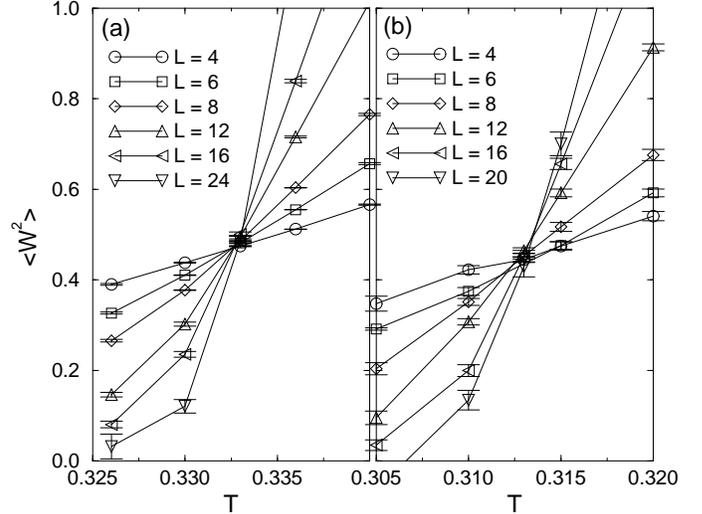}
\caption{ Plot of Monte Carlo data for the the winding number
fluctuations, $\langle W^2\rangle $ vs.\ temperature $T$ for (a) pure
and (b) dirty 3D screened vortex models. According to scaling theory
(see text) the critical temperature is where curves (formed by
straight-line interpolation between data points) intersect.}
\label{winding}
\end{figure}

As noted earlier,
for the purpose of locating the critical point as precisely as
possible for both the pure and random short-range models, we found it
convenient to perform an exact transformation
from the inverted XY model back to the phase representation.  We
compute the spin stiffness $\rho_s$ using the Wolff algorithm to
overcome
the critical slowing down and the difficult problem of equilibrating
different winding number classes.\cite{wolff,moon} From hyperscaling,
$\langle W^2\rangle = L^{d-2}\rho_s/T$ is scale invariant at the
critical point.\cite{rhos} Thus curves for different $L$ all cross at
$T=T_c$ as shown in Fig.\ \ref{winding}.  Using this method, we find
for the pure model $T_c = 0.333\pm0.001$ (which agrees nicely with
the {\em inverse} of $T_c$ for the long range model) and $T_c =
0.313\pm0.001$ for the disordered model.  Of course, this technique
cannot be used to speed up the determination of $z$ since the Wolff
algorithm intentionally does not represent local relaxation dynamics.
Note also that the values of $\langle W^2\rangle $ at the crossing
points in Fig.~\ref{winding} in the pure (a) and disordered (b) cases
agree within the error bars, which is what we expect from two-scale
factor universality\cite{two-scale-factor} and the fact that disorder
is weakly irrelevant to the statics.  

\begin{figure}[htb]
\epsfxsize=\columnwidth
\epsfbox{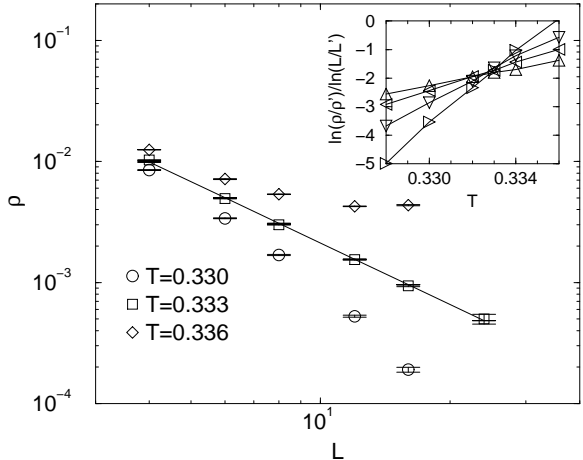}
\caption{Determination of the dynamical critical exponent from MC data
for the linear resistivity $\rho$ computed using the Kubo formula for
the pure screened vortex model.  A power law fit to the data (straight
line) for $T=T_c$ gives the exponent $z = 2.7 \pm 0.1$.  Inset: Data
plotted using the intersection method similarly as in
Fig.~\protect\ref{purel-is_pic}.  The curves have $L/L'=4/6$
(triangles up), $6/8$ (triangles left), $8/12$ (triangles down), and
$12/16$ (triangles right) and give the same values for $z$.}
\label{rho}
\end{figure}

Simulations were then carried out at the measured critical temperatures
to compute the
dynamics in the vortex representations.  In Fig.\ (\ref{rho}) the
dynamical exponent $z$ of the 3D loop model is determined from Monte
Carlo data for the linear resistivity computed from the Kubo formula,
Eq.~(\ref{kubo}).\cite{young94} Using $\rho_{\rm lin} \sim L^{1-z}$
at
$T=T_c$, and from a power law fit to the data we get $z=2.7 \pm 0.1$.
This value is consistent with, but more accurate than, the result of Wengel
and Young\cite{wengel96}.
We also verified that is possible to collapse the data for different
sizes and temperatures using Eq.~(\ref{rho_lin_scale}) with $\nu
\approx {2/3}$.

\begin{figure}[htb]
\epsfxsize=\columnwidth\epsfbox{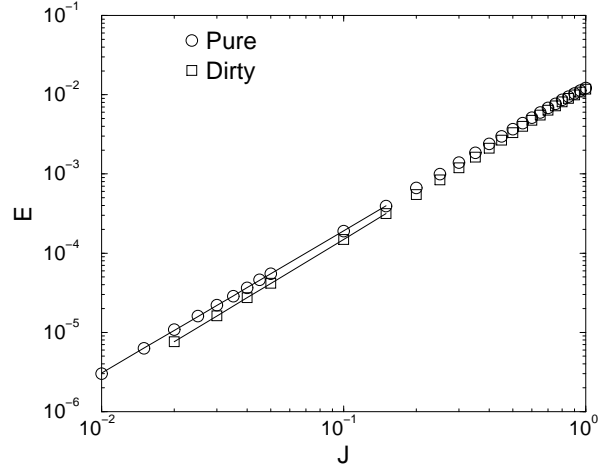}
\caption{Determination of the dynamical critical exponent from MC
data
for the nonlinear I-V characteristics for system size $64^3$.  Power
law fits (solid lines) in the interval of currents where the data is
best described by a power law, with slope $(1+z)/2$ determines $z$.
The dynamical exponents in both cases agree (within the statistical
uncertainty) with the value $z \approx 2.7$ found in
Fig.~\protect\ref{rho}.  }
\label{iv}
\end{figure}

In the case of the disordered vortex model, it is convenient to
determine $z$ from the nonlinear I-V characteristics instead of the
linear resistivity.  Driving the system out of equilibrium makes it
cheaper to converge the simulations, which is desirable for the
disorder averaging.  Fig.~(\ref{iv}) shows data for the nonlinear I-V
characteristics.  For the largest size studied, $L=64$,
it is reasonable to assume that the data is in the
range where $J > J_{\rm nl} \sim 1/L^2$. Hence, according
to Eq.~(\ref{Enonlin}), the dynamical
exponents can be obtained from power law fits of the form $E \sim
J^{(1+z)/2}$ to the data at $T_c$.  The fits were done in the current
interval where a power law best describes the data, leaving out the
highest currents where saturation artifacts in the simulation limit
the voltage.  This gives $z_{\rm pure} = 2.60 \pm 0.1$ and $z_{\rm
dirty} = 2.69 \pm 0.1$.  The small discrepancy between these values
is within the statistical uncertainty of the simulations, and the values
are consistent with the exponent obtained from the calculation of the
linear resistivity.  The coincidence of $z$ for the pure model with
the result of linear resistivity gives a consistency test showing
that the nonlinear I-V characteristics correctly determines $z$.
In particular, our model assumption of a uniform current driving the
vortices in the presence of the disorder does not seem to
introduce any errors.   

\section{Summary and conclusion}

We have performed simulations of
the dynamics of the 3-dimensional XY model in a vortex representation
with and without magnetic screening.  Without screening, we find
$z\approx 3/2$, in agreement with earlier work of Weber and
Jensen\cite{weber97}, who note that this result agrees with the
dynamical critical exponent of the phase model with spin wave degrees
of freedom, model E\cite{hohenberg77}. However, the spin wave degrees
of freedom are {\em separated} from the vortex degrees of freedom when
going to the vortex representation\cite{villain75,jose77,kleinert89}
and the remaining vortex degrees of freedom have only dissipative
motion.  Hence   we find it quite surprising, 
that the exponents from spin wave and vortex dynamics
agree numerically.  We note that there is some experimental evidence for
super-diffusive 3-dimensional XY dynamics in superconductors with
$z\sim 1.25-1.5$, but only for the case of finite magnetic
fields.\cite{ginsberg,moloni-prl}

One can try to argue that it is the long-range forces among the
vortices that accounts for the super-diffusive (i.e. $z<2$) behavior.
Indeed one can even argue that the spin waves are implicitly present
since they are what mediate the long-range forces.  However, we note
that, in this model, the long-range forces are {\em instantaneous}
and
not retarded.  Hence it is still a mystery to us why the vortex model
appears to be consistent numerically with Model E dynamics.
Note that Lee and Stroud,\cite{lee92} who measured $I-V$ characteristics
on the resistively shunted junction model (which is described by
Model E dynamics) 
using Langevin (as opposed to Monte Carlo) dynamics,
find $z=1.5\,(\pm 0.5)$, as expected for this model.

The value of $z$ for the short-range case does not seem to be
significantly affected by the disorder,  and so
disorder appears to be irrelevant dynamically as well as statically.
In both the pure and dirty
case however, $z$ is significantly larger than is usually seen in
relaxational dynamics where $z$ is typically only slightly larger
than
two.\cite{hohenberg77}  There is some experimental evidence for such
enhanced values of $z$.  Anlage's group\cite{3DXY2} finds $\nu=1.0\pm
0.2$, and $z=2.65\pm 0.3$.  The value of $z$ is measured directly at
the critical point from the resistivity scaling and is probably more
reliable than the value of $\nu$ which is measured somewhat more
indirectly (necessarily) using data away from the critical
point.\cite{anlage-private-comm}  It is possible therefore that the
value of $\nu$ is actually consistent with the 3DXY value of 0.667.
Moloni et al.\cite{moloni-prb} find $z=2.3\pm 0.2$.

These experimental values are consistent with the value we obtain here. 
However, in the extreme type-II limit, the inverted XY critical regime 
is expected to be very narrow and difficult to access experimentally.  
It is therefore unclear at this point how significant this agreement
is.  Further work is needed to estimate more precisely
the crossover point to inverted XY behavior in real materials.

To conclude, we have raised a number of issues about experimental and
theoretical uncertainties regarding the dynamical universality class
of the superconducting phase transition in high $T_{\rm c}$
superconductors.  Clearly considerably more work needs to be done to
address this problem.  One aspect that we have not yet
addressed is the effect of disorder in the case of
long-ranged unscreened interactions.  A second problem worth pursuing
is the precise theoretical relationship between the various possible
spin wave dynamics in the phase representation and the corresponding
dynamics of the same model in the various dual (vortex) representations.

\begin{acknowledgements}

C.~W.\ and A.~P.~Y.\ wish to thank Hemant Bokil for useful
discussions
and the Maui High Performance Computing Center for an allocation of
computer time. C.~W.\ and A.~P.~Y.\ are supported by NSF DMR
94-11964. S.~M.~G.\ is supported by DOE MISCON DE-FG02-90ER45427 and
NSF CDA-9601632 and thanks N. Goldenfeld and S. Anlage for
illuminating discussion.  J.~L.\ and M.~W.\ are supported by the Swedish
Natural Science Research Council.  A.~P.~Y.\ and S.~M.~G.\
acknowledge the support of the Aspen Center for Physics.
J.~L.\ and M.~W.\ are supported by the Swedish Natural Science
Research Council, by the Swedish Foundation for Strategic Research
(SSF), and by the Swedish Council for High Performance Computing
(HPDR) and Parallelldatorcentrum (PDC), Royal Institute of
Technology.

\end{acknowledgements}


\begin{references}
\bibitem{global} In principle, global vortex line fluctuations,
extending across the entire sample, are not energetically excluded in
the short range case.  These would change the winding number.
However, since we are only interested in the local time evolution of
the system, we use no such MC moves, thus restricting the MC trial
moves to local plaquette moves also for the vortex model with
screening.
\bibitem{caveat1} This overly-simplified model neglects possible
inhomogeneous current distributions and Amp{\`e}res law.  It is not
known if this affects the critical properties we will consider.
However, note that our result for $z$ for the short range model for
the case of finite currents is consistent with the linear response
result at zero applied current, suggesting that the critical dynamics
is unaffected.
\end{references}
\end{document}